\documentclass[12pt,epsf]{article}
\usepackage{graphicx}
\begin{document}

\newcommand\be{\begin{equation}}
\newcommand\ee{\end{equation}}
\newcommand\ba{\begin{eqnarray}}
\newcommand\ea{\end{eqnarray}}
\newcommand\eq{\begin{equation}}           
\newcommand\en{\end{equation}}
\newcommand\bx{{\bf x}}
\newcommand\delk{\delta_\mathbf{k}}
\newcommand\bu{{\bf u}}
\def\a{\alpha}
\def\b{\beta}
\def\c{\varepsilon}
\def\d{\delta}
\def\e{\epsilon}
\def\f{\phi}
\def\g{\gamma}
\def\h{\theta}
\def\k{\kappa}
\def\l{\lambda}
\def\m{\mu}
\def\n{\nu}
\def\p{\psi}
\def\q{\partial}
\def\r{\rho}
\def\s{\sigma}
\def\t{\tau}
\def\u{\upsilon}
\def\v{\varphi}
\def\w{\omega}
\def\x{\xi}
\def\y{\eta}
\def\z{\zeta}
\def\D{\Delta}
\def\G{\Gamma}
\def\H{\Theta}
\def\L{\Lambda}
\def\F{\Phi}
\def\P{\Psi}
\def\S{\Sigma}

\def\o{\over}
\newcommand{\gsim}{ \mathop{}_{\textstyle \sim}^{\textstyle >} }
\newcommand{\lsim}{ \mathop{}_{\textstyle \sim}^{\textstyle <} }
\newcommand{\vev}[1]{ \left\langle {#1} \right\rangle }
\newcommand{\bra}[1]{ \langle {#1} | }
\newcommand{\ket}[1]{ | {#1} \rangle }
\newcommand{\EV}{ {\rm eV} }
\newcommand{\KEV}{ {\rm keV} }
\newcommand{\MEV}{ {\rm MeV} }
\newcommand{\GEV}{ {\rm GeV} }
\newcommand{\TEV}{ {\rm TeV} }
\def\diag{\mathop{\rm diag}\nolimits}
\def\Spin{\mathop{\rm Spin}}
\def\SO{\mathop{\rm SO}}
\def\O{\mathop{\rm O}}
\def\SU{\mathop{\rm SU}}
\def\U{\mathop{\rm U}}
\def\Sp{\mathop{\rm Sp}}
\def\SL{\mathop{\rm SL}}
\def\tr{\mathop{\rm tr}}

\def\IJMP{Int.~J.~Mod.~Phys. }
\def\MPL{Mod.~Phys.~Lett. }
\def\NP{Nucl.~Phys. }
\def\PL{Phys.~Lett. }
\def\PR{Phys.~Rev. }
\def\PRL{Phys.~Rev.~Lett. }
\def\PTP{Prog.~Theor.~Phys. }
\def\ZP{Z.~Phys. }

\newcommand{\bear}{\begin{array}}  \newcommand{\eear}{\end{array}}
\newcommand{\bea}{\begin{eqnarray}}  \newcommand{\eea}{\end{eqnarray}}
\newcommand{\beq}{\begin{equation}}  \newcommand{\eeq}{\end{equation}}
\newcommand{\bef}{\begin{figure}}  \newcommand{\eef}{\end{figure}}
\newcommand{\bec}{\begin{center}}  \newcommand{\eec}{\end{center}}
\newcommand{\non}{\nonumber}  \newcommand{\eqn}[1]{\beq {#1}\eeq}
\newcommand{\lmk}{\left(}  \newcommand{\rmk}{\right)}
\newcommand{\lkk}{\left[}  \newcommand{\rkk}{\right]}
\newcommand{\lhk}{\left \{ }  \newcommand{\rhk}{\right \} }
\newcommand{\del}{\partial}  \newcommand{\abs}[1]{\vert{#1}\vert}
\newcommand{\vect}[1]{\mbox{\boldmath${#1}$}}
\newcommand{\bib}{\bibitem} \newcommand{\new}{\newblock}
\newcommand{\la}{\left\langle} \newcommand{\ra}{\right\rangle}
\newcommand{\bfx}{{\bf x}} \newcommand{\bfk}{{\bf k}}
\newcommand{\gtilde} {~ \raisebox{-1ex}{$\stackrel{\textstyle >}{\sim}$} ~} 
\newcommand{\ltilde} {~ \raisebox{-1ex}{$\stackrel{\textstyle <}{\sim}$} ~}
\newcommand{\gtrsim}{ \mathop{}_{\textstyle \sim}^{\textstyle >} }
\newcommand{\lesssim}{ \mathop{}_{\textstyle \sim}^{\textstyle <} }
\newcommand{\ds}{\displaystyle}
\newcommand{\bi}{\bibitem}
\newcommand{\lar}{\leftarrow}
\newcommand{\rar}{\rightarrow}
\newcommand{\lrar}{\leftrightarrow}
\def\Frac#1#2{{\displaystyle\frac{#1}{#2}}}
\def\labelenumi{(\roman{enumi})}
\def\SEC#1{Sec.~\ref{#1}}
\def\FIG#1{Fig.~\ref{#1}}
\def\EQ#1{Eq.~(\ref{#1})}
\def\EQS#1{Eqs.~(\ref{#1})}
\def\lrf#1#2{ \left(\frac{#1}{#2}\right)}
\def\lrfp#1#2#3{ \left(\frac{#1}{#2}\right)^{#3}}
\def\GEV#1{10^{#1}{\rm\,GeV}}
\def\MEV#1{10^{#1}{\rm\,MeV}}
\def\KEV#1{10^{#1}{\rm\,keV}}


\baselineskip 0.7cm

\begin{titlepage}

\begin{flushright}
\hfill DESY 06-242\\
\hfill FTPI-MINN-06/41\\
\hfill UMN-TH-2533/06\\
\hfill hep-ph/0612263\\
\hfill December, 2006\\
\end{flushright}

\begin{center}
{\large \bf
The Decay of the Inflaton in No-scale Supergravity
}
\vskip 1.2cm
Motoi Endo${}^{1}$, Kenji Kadota${}^{2}$, Keith A. Olive${}^{2}$, 
Fuminobu Takahashi${}^{1}$ \\
and T. T. Yanagida${}^{3,4}$
\vskip 0.4cm

${}^1${\it Deutsches Elektronen Synchrotron DESY, Notkestrasse 85,\\
22607 Hamburg, Germany}\\

${}^2${\it William I Fine Theoretical Physics Institute, \\
University of Minnesota, Minneapolis, MN 55455, USA}\\

${}^3${\it Department of Physics, University of Tokyo,\\
     Tokyo 113-0033, Japan}

${}^4${\it Research Center for the Early Universe, University of Tokyo,\\
     Tokyo 113-0033, Japan}


\abstract{We study the decay of the inflaton in no-scale supergravity and show that 
decay due to the gravitational interactions through supergravity effects is highly suppressed
relative to the case in minimal supergravity or models with a generic K\"{a}hler potential.
We also show that decay to gravitinos is suppressed. We demonstrate that decay and sufficient reheating are possible with the introduction of a non-trivial gauge kinetic term. This channel may be dominant in no-scale supergravity, yet yields a re-heating temperature which is low enough to avoid the gravitino problem while high enough for Big Bang Nucleosynthesis and baryogenesis. 

}
\end{center}
\end{titlepage}

\renewcommand{\thefootnote}{\alph{footnote}}

\section{Introduction}
\label{int}
Although cosmological inflaton provides an attractive solution to several problems including the horizon/flatness problem \cite{infla,infl}, the realization of an exponentially expanding phase in the early universe in the framework of realistic particle physics models still remains an open question.
As we expect  supersymmetry and supergravity to play a fundamental role in
these models, it is of interest to study inflation in this context as well \cite{susyinfl}.
However, in supergravity models with a minimal K\"ahler potential, all scalar fields generically 
obtain a supergravity mass correction of order the Hubble scale, and the slow-roll condition 
for the inflaton is in general violated, though specific models \cite{simple} can be realized. 
The slow-roll problem found in generic models of supergravity can be avoided if one uses a non-minimal K\"ahler potential of the no-scale form \cite{van}.  Several models of slow-roll inflation in no-scale supergravity have been considered \cite{keithinf,noinf,plan,price}.

No-scale supergravity models based on a non-compact K\"ahler manifold, with a maximally symmetric coset space $SU(N,1)/[SU(N)\times U(1)]$, have attracted substantial interest from string theory because such a K\"ahler potential typically appears in the compactification of higher-dimensional superstring models \cite{comp} as well as from particle physics model building \cite{nosc}.
While local supersymmetry is broken by the no-scale structure of the K\"ahler potential,
there is a residual global supersymmetry leading to 
semi-positive definite scalar potential \cite{van,nosc}. Consequently, one finds that the
tree-level cosmological constant at the global minimum vanishes, in contrast to more generic supergravity models for which the global minimum possesses 
negative vacuum energy density \cite{paul}. 
Indeed, the tree-level potential is flat for the supersymmetry breaking, Polonyi-like field in no-scale models, and so, even though local supersymmetry is broken, the gravitino mass scale is undetermined. From the viewpoint of particle phenomenology, this offers an interesting explanation of the hierarchy problem through the radiative determination of weak/supersymmetry-breaking scale. A non-trivial gauge kinetic function can generate a gaugino mass which breaks global supersymmetry. Radiative corrections set the scale of the gaugino mass, and the form of the gauge kinetic function determines the ratio between the gravitino and gaugino masses.

In addition to the slow-role criteria, another essential feature of
any successful inflationary model is sufficient reheating without the
overproduction of gravitinos.  So long as there is no direct
superpotential coupling of the inflaton to matter fields, i.e. $W =
W(\phi^1) + W(\phi^i)$, where $\phi^1$ is the inflaton and the
$\phi^i$ are matter superfields, the minimal decay rate proceeds via a
3-body gravitational decay with rate $\Gamma \propto (\la \phi^1\ra /M_P)^2 \,
m_{\phi^1}^3/ M_P^2$ \cite{nos,Endo:2006qk}, where $M_P/\sqrt{8 \pi} 
\simeq 2.4 \times 10^{18}\,$GeV denotes the reduced Planck mass.  This
decay leads to a lower limit on the reheat temperature in generic
supergravity models of order $T_{RH} \sim 10^6$ GeV for an inflaton mass
of order $10^{12}$ GeV and an inflaton vacuum expectation value (VEV) 
$\la \phi^1\ra \simeq M_P$. While the thermal production of gravitinos at
this temperature is not problematic, the direct decay of the inflaton
imposes strong constraints on inflation models
\cite{KTY,grasca}. These results also hold in large field
inflation models such as chaotic~\cite{Kawasaki:2000yn} and
hybrid~\cite{Copeland:1994vg} inflationary models.

In this letter, we will study the corresponding questions of reheating and 
gravitino production in no-scale supergravity models.  In fact, we find that the decay of the inflaton is highly suppressed in no-scale supergravity, which is directly related to the special structure of the 
K\"ahler manifold in no-scale models. 
In particular, the tree-level gravitational interactions of the inflaton due to supergravity effects  
vanish exactly at the global minimum. 
As a consequence, without the direct coupling of the inflaton to matter
(which may be problematic for model building) the inflaton is stable at the tree-level.
Therefore, the suppression of the gravitational interactions of the inflaton due to the symmetry of
 K\"ahler manifold is of prime importance
in inflation model building in supergravity.

We show, however, that the introduction of a  non-trivial gauge kinetic function can lead to 
the decay of the inflaton with successful reheating~\cite{Endo:2006tf} 
but without the overproduction of gravitinos. 
Because the dominant decay channel is specified by the gauge kinetic function 
in no-scale models, we can constrain the 
gauge kinetic function by the reheating temperate constraints. 
Interestingly, since the gauge kinetic function relates the gravitino
and gaugino masses, the reheating temperature constraints
provide an upper bound on the gravitino mass.

\section{Setup: No-scale Supergravity}
\label{setup}
We consider a no-scale model with a K\"ahler
potential of the form~\cite{van,nosc}\footnote{
We require that the argument
of the logarithmic function be positive since otherwise the kinetic terms of the matter fields have wrong
signs, or equivalently,  unitarity is broken.
}
:
\bea
\label{eq:no-scale}
K &=& -3 \ln\left[z+z^\dag -\frac{1}{3}\left(  \sum_{i=1}
|\phi^i|^2 \right) \right]
\eea 
with a supersymmetry breaking field $z$, an inflaton $\phi^1$, matter fields $\phi^i(i=2,3, ...)$ and as
noted earlier, the superpotential is assumed to include no direct coupling between an inflaton and the other fields, $W=W(\phi^1)+W(\phi^i)$.
We also assume that the superpotential does not contain $z$, so that the tree-level potential for $z$ remains flat at the minimum (one of the notable features in a no-scale supergravity model). 
Unless explicitly noted, we will adopt Planck units so that $M_P/\sqrt{8\pi} = 1$. The total K\"ahler potential $G$ is defined as $G \equiv K + F + F^\dagger$ with $F\equiv \ln W$.

The relevant bosonic kinetic terms are derived from
$G_a^b (\partial_\mu \phi^a) (\partial^\mu \phi_b^*) = 
K_a^b (\partial_\mu \phi^a) (\partial^\mu \phi_b^*)$ for all scalar fields
including $z$, the inflaton,  and matter. 
Indices on the K\"ahler potential refer to derivatives with respect to the fields, 
$G_a = \partial G/\partial \phi^a$, $G^b = \partial G/ \partial \phi_b^*$, etc.  
After some algebraic rearrangements, these can be written as
\ba
\label{boskin}
\frac{1}{12}(\partial_{\mu} K)^2+e^{K/3}|\partial_{\mu}\phi^i|^2 
-\frac34 e^{2K/3}|\partial_{\mu}(z-z^{*})
+\frac13(\phi_i^{*}\partial_{\mu}\phi^i-\phi^i\partial_{\mu}\phi^{*}_i)|^2
\ea 
The vector field kinetic terms are specified by an additional function $f_{\alpha \beta}$ as 
\ba
-\frac14(Ref_{\alpha \beta})(F_{\alpha})_{\mu \nu}F^{\mu \nu}_{\beta}
\ea
and the scalar potential is given as
\ba
\label{scapot}
V=e^G[G_i(G^{-1})^i_jG^j-3]+\frac12 Ref_{\alpha \beta}^{-1}D^{\alpha}D^{\beta}=e^G e^{-K/3} F_i F^{i \dagger}+\frac12 Ref_{\alpha \beta}^{-1}D^{\alpha}D^{\beta}
\ea 
where the D-term is given by $D^{\alpha}=gG^iT^{\alpha j}_{i}\phi_j$ with a gauge coupling constant $g$ and the generator of the gauge group $T^{\alpha}$. Derivatives of the superpotential are denoted by  $F_i= W_i/W = \partial \ln W/\partial \phi^i$.
It should be noted that there are no soft masses for scalar fields 
in the scalar potential. Further, it is known that even anomaly-mediated 
SUSY breaking effects vanish in no-scale supergravity models~\cite{Luty:1999cz}.
This feature enables us to have an arbitrarily large gravitino mass by choosing an 
appropriate  gauge kinetic function~\cite{plan}.  We will come back to this point in
Sec.~\ref{sec:5}.

The theory is completely defined once $G$ and the gauge kinetic function $f_{\alpha\beta}$ are specified. 
For now, we will take $f_{\alpha\beta} = \delta_{\alpha\beta}$.  We will consider a non-trivial form
for the gauge kinetic function in Sec. \ref{gaugesec} when we 
discuss the possible inflaton decay channel through the terms involving this function.

As one can see from Eq. (\ref{scapot}), the scalar potential takes a form reminiscent of 
globally supersymmetric models.  Indeed, it can be rewritten as
\ba
V =  e^{2K/3} W_i W^{i \dagger}+\frac12 \delta_{\alpha \beta} D^{\alpha}D^{\beta}
\label{scapot2}
\ea
The above scalar potential is semi-positive definite and, from now on, we assume 
$<F^i>=<F_i^{\dagger}>=<D^{\alpha}>=0$ at the minimum to ensure the vanishing of the cosmological constant.

\section{Inflaton mass eigenstate}
\label{inflamass}

From the form of the K\"ahler potential and the scalar kinetic terms in Eq. (\ref{boskin}), it is clear that we
have defined the theory in a basis with non-minimal kinetic terms.
In discussing the decay of an inflaton field, it will be useful to define
the inflaton mass eigenstate in a basis with canonically normalized fields.
The canonically normalized scalar fields can be read off from 
Eq. (\ref{boskin}) 
\ba
Z_R&=&\sqrt{\frac16}K
\label{zr}
\\
i Z_I&=& e^{<K>/3}\sqrt{\frac32}(z-z^*+\frac 13 \phi_{01}^*\delta \phi^1-\frac13\phi_0^1\delta \phi_1^*)
\\
\label{scalar}\Phi^i&=&e^{<K>/6}\phi^i
\\
A_\mu^{'\alpha}&=&<Ref_{\alpha \beta}>^{1/2}A_\mu^{\beta}
\ea
where $A_\mu$ is a gauge boson and we assumed that the scalar components of $z$ and $\phi^1$ have finite vacuum expectation values (VEVs),
$z_0$ and $\phi^1_0$ respectively ($\delta \phi^1$ is the fluctuation around $\phi^1_0$).
We assume that all of the other  $\phi^i$ have vanishing VEVs due to some symmetries (e.g. gauge symmetries) \footnote{The general procedure to obtain the
mass-eigenstate basis was presented in Ref.~\cite{Endo:2006tf}.}.

It is straight forward to calculate the scalar mass matrix in terms of these canonically normalized fields. The scalar mass matrix elements involving the canonically normalized inflaton field are
\ba
<\frac{\partial^2 V}{\partial \Phi^1 \partial \Phi_j^*}> &=& <\frac{\partial \phi^1 }{\partial \Phi^1}\frac{\partial \phi^*_j}{\partial \Phi^*_j}\frac{\partial ^2 V}{\partial \phi^1 \partial\phi^*_j }>
\nonumber \\
 &=&
e^{<G>}e^{<-2K/3>}<F_{1k}{F^{\dagger}}^{kj}>
\label{smass}
\ea
where we have used $\langle F_i \rangle = 0$ for all i, and $\langle K_i \rangle = 0$ for $i \ne 1$
(note that subscripts indicating the derivatives are with respect to the model (un-normalized) fields).
This leads to 
\ba
\langle\frac{\partial^2 V}{\partial \Phi^1 \partial \Phi^*_j}\rangle=e^{<K/3>}
<W_{1k}{W^*}^{kj}>
\ea
which vanishes in the absence of direct coupling terms between an inflaton and the other fields in the superpotential, $<W_{1j}>=0$ for $j \ne 1$. Note that we also have
\ba
<\frac{\partial^2 V}{\partial \Phi_1^* \partial \Phi_j^*}>=<\frac{\partial^2 V}{\partial \Phi^1 \partial \Phi^{j}}>=
<\frac{\partial^2 V}{\partial \Phi_1^* \partial Z_{R,I}}>=<\frac{\partial^2 V}{\partial \Phi^1 \partial Z_{R,I}^*}>=0
\ea 
which are obtained from $<F^{i}>=<F_{i}^*>=0$ at the minimum. 

Hence, if there is no direct coupling between the inflaton and the other fields in $W$, 
the canonically normalized inflaton field is the 
inflaton mass eigenstate.

Before starting the discussion for the suppression of the decay of $\Phi^1$, we make a brief comment on an
approximate symmetry of $\delta \Phi^1$ (the fluctuation around its VEV $<\Phi^1>$). In terms of the canonically normalized fields, the dependence of $G$
 on the inflaton $\Phi^1$ appears only in the superpotential. Since the linear term of $\delta \Phi^1$ 
in the superpotential should vanish from $<\partial W/\partial \Phi^1>=
<\partial W/\partial \delta \Phi^1>=0$, the lowest order term in $W$ is quadratic in $\delta \Phi^1$. Noting that the higher order terms in $\delta \Phi^1$ do not affect the decay of the inflaton, $G$ has an approximate $Z_2$ symmetry of $\delta \Phi^1$. Taking this into account, one can guess that spontaneous inflaton decay 
should be suppressed (apart from possible $Z_2$ symmetry breaking via terms including the gauge kinetic function $f_{\alpha \beta}$). 
In the following, we show explicitly that inflaton decay is indeed suppressed for the terms which are solely determined from $G$ 
in Sec. \ref{fersec}-\ref{grasec}, followed by the discussion of the possible inflaton decay via terms involving $f_{\alpha \beta}$ in Sec. \ref{gaugesec}.

\subsection{Inflaton coupling terms from the mass matrix expansion}
\label{fersec}
We can consider the expansion of the mass matrix to study the possible decay of $\Phi^1$.
For example, the scalar mass matrix $\Phi^*_i ({\cal M}_0^2)^i_{j} \Phi^j (i,j\ne 1)$ can 
in principle give $\Phi^1$ coupling terms such as
$<\partial    ({\cal M}_0^2)^i_{j}/\partial \Phi^1>\delta \Phi^1 \Phi^*_i \Phi^j$. Such terms however vanish due to the special form of K\"ahler 
potential in a no-scale model because
\ba
\left<\frac{\partial}   {\partial \Phi^1}  ({\cal M}_0^2)^i_{j}\right> \delta \Phi^1 \Phi^*_i \Phi^j
&=&\left<\frac{\partial}   {\partial \Phi^1}  e^G e^{-2K/3}F_{jk}{F^\dagger}^{ki}\right> \delta \Phi^1 \Phi^*_i \Phi^j \nonumber \\
&=&
\left<\frac{\partial}   {\partial \Phi^1}  e^{\sqrt{2/3}Z_R}W_{jk}{W^*}^{ki}\right> 
\delta \Phi^1 \Phi^*_i \Phi^j \nonumber \\
\nonumber \\
&=&\left< e^{\sqrt{2/3}Z_R} \frac{\partial \phi^1}{\partial \Phi^1}\frac{\partial}{\partial \phi^1}  W_{jk}{W^*}^{ki}\right> \delta \Phi^1 \Phi^*_i \Phi^j \nonumber \\
&=&0
\ea 
where the last equality is due to the assumption that there are no terms which directly couple 
an inflaton to the other fields in $W$.
In the above, we also used the fact that $<\partial D^{\alpha i}/ \partial \Phi^1>=0$, so that the D-term does not contribute to inflaton decay either.
One also finds that inflaton decay coming from the expansion of the other scalar mass terms vanish by performing 
the analogous calculations as above, such as $<\partial ({\cal M}_0^2)_{Z_R j}/ \partial \Phi^1>=0$. 
Note that the absence of a decay term for the inflaton  to scalars is a direct result of the no-scale
form of the potential in Eqs. (\ref{scapot}) and (\ref{scapot2}).

Similarly, the expansion of matter Fermion mass matrix terms $\bar{\chi^{i}}{{\cal M}_{1/2}}_{ij}{\chi^{j}}$ as well as the matter Fermion-gaugino mass matrix  $\bar{\lambda}_{\alpha}{{\cal M}_{1/2}}^{\alpha}_i{\chi^{i}}$  give a vanishing contribution for inflaton decay. 
In general, we can write the chiral fermion mass matrix as
\ba
\bar{{\chi}^{i}}{{\cal M}_{1/2}}_{ij}{\chi^{j}}
=\- e^{G/2}\bar{\chi^{i}}\left(G_{ij}+G_{i}G_{j}-G_{ij}^{m}(G^{-1})_{m}^{n}G_{n}\right){\chi^j}
\ea
For the specific case of no-scale supergravity, this can be rewritten as \footnote{The following arguments are not affected even if one uses the canonically normalized Fermion fields $\chi^{'i}= e^{<K>/6}\chi^{i},\lambda^{'\alpha}=<Ref_{\alpha \beta}>^{1/2} \lambda ^{\beta}$.}
\ba
- e^{G/2}\bar{\chi^{i}}\left(\frac23G_{i}G_{j}+F_{ij}+\frac13F_{i}F_{j}\right){\chi^{j}}
\ea
Subtracting the Goldstino component $\eta=(G_i)\bar{\chi}^i$ in the unitary gauge, the Fermion mass matrix term becomes
\ba
-e^{K/2}\bar{\chi^i} \left(W_{ij}-\frac{2}{3}\frac{W_iW_j}{W}\right) {\chi^j}
\label{yukawat}
\ea
Any possible inflaton decay channels to two chiral fermions can be obtained by the expansion of Eq. (\ref{yukawat}) around the VEV of $<\Phi^1>$. As one can see
\ba
&&-\left<\frac{\partial}{\partial \Phi^1} e^{-\sqrt{3/2}Z_R}
\left(W_{ij}-\frac{2}{3}\frac{W_iW_j}{W}\right) \right>   \delta \Phi^1   \bar{\chi^i} {\chi^j}
\nonumber \\
&=&-\left<e^{-\sqrt{3/2}Z_R} \frac{\partial \phi^1}{\partial  \Phi^1}
\frac{\partial}{\partial \phi^1}
\left(W_{ij}-\frac{2}{3}\frac{W_iW_j}{W}\right) \right>    \delta \Phi^1   \bar{\chi^i} {\chi^j}
\ea
and
\ba
&&-\left<\frac{\partial}{\partial \Phi^k}\frac{\partial}{\partial \Phi^1} e^{-\sqrt{3/2}Z_R}
\left(W_{ij}-\frac{2}{3}\frac{W_iW_j}{W}\right) \right>   \delta \Phi^1  \Phi^k   \bar{\chi^i} {\chi^j}
\ea
vanish because $<W_i>=< W_1 >=0$ and we have assumed no direct coupling between inflaton and the other fields in $W$. Indeed, we can easily see from the above procedures that all the inflaton decay channels due to the expansion of the Fermion mass terms vanish at the tree level.
%

For completeness, we also consider the inflaton coupling to gauginos and matter Fermions, 
\ba
2igG^i_j(T^{\alpha})_{ik}\phi^k\bar{\lambda}_{\alpha}\chi^{i}
\ea
Inflaton decay from this term also vanishes because 
$\left<{\partial G^i_j}/{\partial \Phi^1}\right>=0$.

Hence all tree-level  inflaton decays to  scalar and fermion matter fields including gauginos  exactly vanish at the global minimum in no-scale supergravity models. 

\subsection{Derivative coupling}
\label{kinsec}
Next we consider possible inflaton decays via kinetic terms.
From Eq. (\ref{boskin}), we see that the inflaton, $\phi^1$, may
couple to scalar fields $\phi^i$ through, 
\beq
\label{eq:matter-th-K}
 - \la \frac{\partial e^{K/3}}{\partial \phi^1} \ra\, \delta \phi^1 \del  \phi^i\del \phi_i^\dag + {\rm h.c.}
\eeq
where again $\delta \phi_1$ is the fluctuation around its VEV. However, we can easily see that, by considering the canonically normalized 
scalar fields, the total decay of the inflaton mass eigenstate 
should vanish. All of the derivatives of the coefficients in the kinetic terms in Eq. (\ref{boskin}) with respect to $\Phi^1$ vanish because $<\partial Z_R/\partial \Phi^1>=0$. Hence the $\Phi^1$ decay channel
to canonically normalized scalar fields which can be obtained by the expansion of the 
kinetic terms in Eq. (\ref{boskin}) exactly vanish. Couplings such as those in Eq. (\ref{eq:matter-th-K})
are absorbed by the canonically normalized mass-eigenstate field $Z_R\equiv K/\sqrt6$.
The same is true for the matter Fermion kinetic terms
\ba
-e^{K/3}\sum_{i=1}^{n-1}\bar{\chi}_i \not\! {\partial}\chi^i 
\ea
Thus the inflaton mass-eigenstate $\Phi^1$ does not decay through the kinetic terms.

\subsection{Gravitinos}
\label{grasec}
Finally, let us now consider the decay of the inflaton into a pair of gravitinos. 
Gravitino overproduction from inflaton decay is quite dangerous
because the decay of  gravitinos may affect the light-element abundances 
or the density of the  lightest supersymmetric particles (LSPs) produced by
gravitino decay may exceed the allowed dark matter abundance. 
The relevant interactions for gravitino pair production are 
given by~\cite{WessBagger}
\bea
   - \frac{1}{8} \epsilon^{\mu\nu\rho\sigma}
   \left( \frac{\partial G}{\partial \Phi^1} \partial_\rho \Phi^1 -  
\frac{\partial G}{\partial \Phi_1^* }\del_\rho \Phi_1^*
     \right)
   \bar \psi_\mu \gamma_\nu \psi_\sigma 
   - \frac{1}{8} e^{G/2} \left(\frac{\partial G}{\partial \Phi^1} \delta \Phi^1  +
 \frac{\partial G}{\partial \Phi^*_1} \delta \Phi_1^*
    \right)
   \bar\psi_\mu \left[\gamma^\mu,\gamma^\nu\right] \psi_\nu
   \label{eq:phi2gravitino}
\eea
where $\psi_\mu$ is the gravitino field, and we have chosen the
unitary gauge in the Einstein frame. Thus the inflaton couplings with 
gravitinos are proportional to $\la \partial G/\partial \Phi^1 \ra = 0$. 
Once again, for completeness, we display the scalar-gravitino-gaugino coupling
\ba
 -\frac{i}{2}gG^i(T^{\alpha})_{ij}\phi^j\bar{\psi}_{\mu}\gamma^{\mu}\lambda_{\alpha}
\ea
which also can not contribute to inflaton decay because
$\left<{\partial G^i}/{\partial \Phi^1}\right> = 0$. Therefore, gravitino pair production rate from 
inflaton decay vanishes exactly at the tree-level.

\section{Inflaton decay via gauge kinetic function}
\label{gaugesec}

The exercise of the previous section shows that in the context of no-scale supergravity, 
not only is there no gravitino problem due to excess reheating, there is virtually no reheating at all.
Thus we are presented with a potentially more severe problem for 
inflation in no-scale supergravity.
Finding that the inflaton decay in no-scale supergravity model is indeed highly suppressed, the natural question now would be to 
find the dominant decay channel of an inflaton to reheat the universe. 

The absence of supersymmetry breaking scalar masses in Eq. (\ref{scapot}) is one of the features of a no-scale model and one mechanism to mediate supersymmetry breaking to the visible sector can be specified through a z-dependent gauge kinetic function. In this section, we show that 
 terms involving a non-trivial gauge kinetic function can be responsible for the dominant  
 channel of $\Phi^1$ decay~\cite{Endo:2006tf}.

Among the terms involving the gauge kinetic function, the terms of interest here are 
\ba
\label{gauge}
-\frac14 (Ref_{\alpha \beta})F^{\alpha}_{\mu \nu}F^{\beta \mu\nu}
+\frac{i}{4}(Im f_{\alpha \beta})\epsilon^{\mu\nu\rho\sigma}F^{\alpha}_{\mu \nu}F^{\beta \rho\sigma}
+\left(\frac14e^{G/2}\frac{\partial f_{\alpha\beta}^*}{\partial \phi^{*}_j}(G^{-1})_j^kG_k\lambda^{\alpha}\lambda^{\beta}+h.c.\right) 
\label{gaugino}
\ea
Other terms involving the gauge kinetic function are the derivative coupling terms, and the inflaton decay rate from those terms are suppressed by the masses of the final state particles.
For illustrative purposes, let us take the simplest non-trivial form for $f_{\alpha\beta}$
such that it depends only on the field $z$
\ba
f_{\alpha\beta}\equiv \delta_{\alpha\beta}h(z)
\ea
This simple choice determines the universal Majorana (canonically normalized) gaugino mass at the unification scale for softly broken global supersymmetry
\ba
m_{1/2} = \left|\frac12  e^{G/2} \frac{h_z}{Re~h} (G^{-1})_z^kG_k\right|=
\left|\frac12  e^{(G/2-K/3)}  \frac{h_z}{Re~h} (1 - \phi F_{\phi \phi})\right|
\label{gmass}
\ea
where $h_z\equiv \partial h/\partial z$ and we used the relation for
the projection operator $\alpha_z = (G^{-1})_z^kG_k = -e^{-K/3}(1 - \phi F_\phi/3)$\footnote{Note the gaugino mass
term is related to the gravitino mass $m_{3/2}=|e^{G/2}|$, but, in general, 
the exact ratio can vary depending on the form of $f_{\alpha\beta}$, cf. Eq. (\ref{gmass}).}.

The decay of the inflaton to two gauge bosons can be obtained from the expansion of Eq. (\ref{gauge})
\ba
-\frac14 \left<\frac{\partial}{\partial \Phi^1} h \right>  \delta \Phi^1  F_{\alpha \mu \nu}F^{\alpha \mu\nu} &=& 
-\frac14 \left< h_z \frac{\partial z}{\partial \Phi^1}  \right>  \delta \Phi^1  F_{\alpha \mu \nu}F^{\alpha \mu\nu}  \nonumber \\
&=& -\frac1{12} \left< h_z e^{-K/3} \Phi^{*}_1  \right>  \delta \Phi^1  F_{\alpha \mu \nu}F^{\alpha \mu\nu} 
\ea
where we have used $\partial z/\partial \Phi^1 = e^{-<K/3>} \Phi^*_1/3$.
Thus, the inflaton decay rate to canonically normalized gauge bosons from this coupling 
becomes of order \footnote{For the numerical estimation, we assumed there are a total of 12 gauge bosons as in MSSM.}  
\ba
\Gamma(\Phi_1\rightarrow A_\mu A_\mu) \sim {\cal O}(10^{-3})\times 
e^{-\frac{<K>}{3}}\left|\left<\frac{ e^{-K/6}\Phi_1}{M_p}\right>\right|^2\left|\left<\frac{h_z}{Re~h}\right>\right|^2  \frac{m_{\phi^1}^3}{M_p^2}
\label{grate}
\ea

In addition, there is also a non-negligible contribution for the inflaton 
decay to gauginos via the gaugino mass term in Eq. (\ref{gaugino}) 
which from Eq. (\ref{gmass}) becomes of order 
\ba
\frac14 \left< e^{G/2-K/3} \frac{\partial}{\partial \Phi^1} \left[ \frac{h_z}{Re~h} 
(1 - \phi^1 F_1 / 3) \right] \right>  \delta \Phi^1 \lambda^{\alpha'} \lambda^{\alpha'} \nonumber \\
= \frac1{12} \left< e^{G/2-2K/3} \left[ \left( \frac{h_{z}}{Re~h} \right)_z  \Phi^*_1   - 
\frac{h_z}{Re~h} \Phi^1 F_{11} \right]
 \right>  \delta \Phi^1 \lambda^{\alpha'} \lambda^{\alpha'}
 \label{ginocoup}
\ea
where $\lambda'$ is the canonically normalized gaugino field
$\lambda'^{\alpha}=<Re~h>^{1/2}\lambda^{\alpha}$.
If $(h_z/Re h)_z$ is small, the second term in Eq. (\ref{ginocoup}) dominates and we obtain a decay rate to gauginos
\ba
\Gamma(\Phi_1\rightarrow {\lambda'}\lambda')\sim {\cal O}(10^{-3})\times
e^{-\frac{<K>}{3}}\left|\left<\frac{ e^{-K/6}\Phi^1}{M_p}\right>
\right|^2\left|\left<\frac{h_z}{Re~h}\right>\right|^2  \frac{m_{\phi^1}^3}{M_p^2}
\ea
as in Eq. (\ref{grate}), where we have used $m_{\phi^1} = |e^{G/2-K/3}F_{11}|$ 
(from Eq.(\ref{smass})). 

The reheating temperature 
in our example is estimated to be of order
\ba
\label{eq:trh}
T_{RH}\sim {\cal O}(10^7) \times \left|\left<\frac{h_z }{(Re~h)}\right>\right| {\rm\, GeV}
\ea
for $m_{\phi^1} \sim 10^{-7}$ (coming from the constraints on the 
cosmic perturbation amplitude~\footnote{We are considering here small field inflation models, such
as that in \cite{keithinf}, with an inflaton VEV
$<\phi>\sim{\cal O}(1)$. In fact, inflation models with a VEV $<\phi> \ll{\cal O}(1)$ would suffer
from insufficient reheating in no-scale supergravity.})
and $\la \Phi^1 \ra \simeq M_P$, which is low enough to avoid the gravitino problem \cite{bbn} unless $<{h_z}/{(Re~h)}>$ is tuned to be much larger than ${\cal O}(1)$.


\section{Discussion and conclusion}
\label{sec:5}
In this letter we have shown that the many inflaton decay channels to the matter fields and gravitinos are highly suppressed when the K\"ahler potential is of the no-scale form, and that the dominant 
inflaton decay channels depend on the gauge kinetic function. 
As a consequence, the decay of the inflaton connects the reheating temperature
with the relation between gaugino and gravitino masses.
For example, in our example, the reheating temperature is proportional to the
ratio of the gaugino mass to the gravitino mass determined by $<h_z/Re~h>$. This is in contrast to the discussions for the other inflation models in
no-scale supergravity previously considered \cite{noinf,plan} where either explicit inflaton couplings or non-trivial
gravitational couplings are assumed for the inflaton decay. The direct coupling of an inflaton and other fields in superpotential, for instance, will spoil the flatness of the inflaton potential \footnote{For example, it may be hard to 
prevent the couplings via non-renormalizable operators. The inclusion of a non-renormalizable coupling such as $W\ni \lambda_n \phi^1 {\phi^i}^n$ ($n>2$) in addition to the superpotential $W\ni \mu^2 (\phi^1-(\phi^{1})^4/4)$ which can lead to inflation in no-scale supergravity \cite{keithinf} would require the fine-tuning of the coupling constant $\lambda_n \ll 10^{-7}$ to insure the flatness of the inflaton potential if the $\phi^i$  obtain a large (say Planck scale) vev due to the quantum fluctuations during inflation. Even if the $\phi^i$ happen to have the vanishing vev during inflation so that the inflaton potential around the origin is not affected by such a coupling, we may still derive a limit from the reheating temperature due to the additional decay channels of the inflaton which is of order $\lambda_n \lesssim 10^{-21+7n}$ (assuming the conservative limit of 
$T_{RH} < 10^8 GeV$).} even though no-scale K\"ahler potential can still help suppressing the reheating temperature and the gravitino production. 

The low reheating temperature certainly helps resolving the gravitino problem. We note, however, that the gravitino problem is not generic in no-scale supergravity 
models, as the gravitino mass is a priori undetermined and may be as large as the 
Planck scale~\cite{plan}. As we have discussed, in generic inflation models based on no-scale supergravity, a Planck mass 
gravitino would require a very small value for $h_z$ restoring our initial problem of 
sufficient reheating. For instance, if we require a reheating temperature $T_{RH}$
larger than $10{\rm\, MeV}$~\cite{sw,Kawasaki:1999na,Hannestad:2004px},
$<h_z/Re~h>$ cannot be smaller than ${\cal O}(10^{-9})$ from Eq.~(\ref{eq:trh}),
which leads to $m_{3/2} \lesssim {\cal O}(10^{12}) {\rm\,GeV}$ for 
$m_{1/2} = {\cal O}(10^3) {\rm\,GeV}$ (assuming $e^{-K/3} \sim {\cal O}(1)$,
see Eq.~(\ref{gmass})). 

Although the non-minimal couplings
between the inflaton $\phi^1$ and the supersymmetry breaking field $z$ induce gravitino overproduction in a generic supergravity model,
as we saw in Sec.~\ref{grasec},  inflaton decay into a pair of gravitinos
is suppressed in no-scale supergravity . This suppression is quite important.  
If such suppression did not occur,
it would be very difficult to satisfy the BBN constraints~\cite{bbn} for an unstable gravitino of a mass
$m_{3/2} = 100{\rm\,GeV} - 10{\rm\,TeV}$. Further, even for gravitinos heavier than
$10{\rm\,TeV}$, the abundance of the LSPs produced by gravitino decay
might exceed the dark matter abundance \footnote{For gravitinos
heavier than ${\cal O}(10^{3-4})\,$TeV, the resulting temperature after 
gravitino decay 
can be high enough for LSPs 
to annihilate efficiently (or reach thermal equilibrium). Then the overclosure 
problem can be resolved.}. In particular,  as long as the decay via the gauge kinetic function is the dominant source for reheating,
the reheating temperature is inversely proportional to the gravitino mass. 
So, heavier gravitinos and a lower reheating temperature as a 
result of a smaller decay rate of the inflaton via the gauge kinetic function, 
would lead to a larger branching ratio for gravitino 
production~\cite{KTY}, making the problem more severe.  Fortunately, gravitino production
is suppressed in no-scale supergravity, and the problems mentioned above are avoided.

The production of a baryon asymmetry is also somewhat constrained. For example, baryo/leptogenesis through
out-of-equilibrium decay normally requires the inflaton to decay directly into fields
generating the asymmetry. In the no-scale models discussed above, the inflaton 
decays directly only into gauge bosons and gauginos, thus severely limiting possible 
mechanisms. Models such as the Affleck-Dine mechanism at low temperature
would remain plausible possibilities \cite{ad}.   

As mentioned before, since the modulus $z$ has a flat potential at the tree level,
there is a moduli problem associated with the $z$ field.
The moduli problem \cite{mod} is a prevailing problem in many inflation models, 
and is quite often a more serious problem than the gravitino problem. 
Firstly, one needs to stabilize the moduli. 
For instance, to avoid the run-away minimum during inflation, 
one may need to modify the K\"ahler potential \cite{plan} or add additional D-term or non-perturbative effects \cite{price}. 
Secondly, even if we can stabilize the moduli, one still needs to worry about their late-time decay which can jeopardize Big Bang Nucleosynthesis \cite{bbn}. 
Furthermore, it should be noted that, once $z$ is stabilized by e.g. introducing 
non-trivial $z$-dependence of the superpotential or modifying the
structure of the K\"ahler potential,  the anomaly-mediation effects are generically 
non-negligible. Then, it would be difficult to have the gravitino mass larger than ${\cal O}(100)\,$TeV 
on the basis of naturalness. 

Finally, we note that, analogous to the discussion of the decay of $\Phi^1$ to gravitinos in Sec. \ref{grasec}, the modulus $Z_R$ (which is also a mass eigenstate~\footnote{This may not be the case if $z$ is stabilized by modifying its potential, which, however, does not essentially affect the following discussion.})
couples to gravitinos with a coupling of order $e^{G/2}G_{Z_R}/8$. Recalling that $<G_{Z_R}>=\sqrt6$, 
the modulus decay to the gravitinos may not be negligible (the possible significance of the modulus decay to the gravitinos was also pointed out in 
\cite{mod1}), if the decay is kinematically allowed. 
One possibility to avoid this problem of late-time moduli 
decay is the enhancement of the moduli decay at the minimum \cite{enh}.  
We leave the study of the moduli problem in 
no-scale supergravity for the future work. 

\noindent{ {\bf Acknowledgments} } \\
\noindent
K.K. and K.A.O would like to thank Tony Gherghetta and John Ellis for useful discussions. The work of K.K. and K.A.O. was partially supported by DOE grant DE--FG02--94ER--40823.

\end{document}